\documentclass[10pt,letterpaper,twocolumn]{article} 

\usepackage{ol2}
\usepackage[draft]{hyperref}
\usepackage{amsmath}

\begin{document}

\twocolumn[ 

\title{Aharonov-Bohm photonic cages in waveguide and coupled resonator lattices by synthetic magnetic fields}


\author{Stefano Longhi}

\address{Dipartimento di Fisica, Politecnico di Milano and Istituto di Fotonica e Nanotecnologie del Consiglio Nazionale delle Ricerche, Piazza L. da Vinci 32, I-20133 Milano, Italy (stefano.longhi@polimi.it)}

\begin{abstract}
We suggest a method for trapping photons in quasi one-dimensional waveguide or coupled-resonator lattices, which is based on an optical analogue of the Aharonov-Bohm cages for charged particles. Light trapping results from a destructive interference of Aharonov-Bohm type induced by a synthetic magnetic field, which is realized by periodic modulation of the waveguide/resonator propagation constants/resonances. 
\end{abstract}



 ] 

\noindent 
Realization of artificial magnetic fields at small length scales in photonic structures has received a great and increasing interest in the past few years, paving the way toward the
realization of novel mechanisms for the control and manipulation of light \cite{M1,M2,M2bis,M3,M4,M5,ref2,M6,M7,M8,M8bis,M9}. Photons subjected to artificial gauge fields can behave like electrons in topological insulators \cite{ref1}, thus sustaining topologically-protected edge modes that propagate immune of disorder \cite{M3,M4,M8,M9}. Like for charged particles in a magnetic field, photons in an artificial gauge field also acquire an additional phase of the Aharonov-Bohm (AB) type, which can be exploited to realize  AB interference effects with photons. AB effects for photons have been predicted in several theoretical works (see, e.g., \cite{AB0,AB1,AB2,AB3}) and observed in a few recent experiments \cite{AB4,AB5,AB6} at both microwave and optical frequencies. \par
In this Letter we suggest a method for trapping photons in quasi one-dimensional waveguide or coupled resonator lattices, which is inspired by a photonic analogue of AB caging for charged particles \cite{Cage1,Cage2,Cage3,Cage4,Cage5,Cage6,Cage7,Cage8}.  AB caging, originally proposed in Ref.\cite{Cage1} for tight-binding electrons
in certain two-dimensional lattices (including rhombic chains and $\mathcal{T}_3$, $\mathcal{T}_4$ periodic tilings),  exploits the destructive interference among different hopping paths at some special values of the magnetic flux.  Such a destructive interference bounds the set of sites that can be visited by an initially localized wave packet, as schematically shown in Fig.1(a) for a rhombic chain. This kind of localization has been demonstrated in superconducting wire networks \cite{Cage5}, mesoscopic semiconductor lattices \cite{Cage6} and arrays of
Josephson junctions \cite{Cage7}, but not yet for photons. We note that AB caging is very distinct from other forms of localization in ordered lattices, such as dynamic localization and coherent destruction of tunneling, which have been demonstrated for light beams in recent experiments \cite{CDT1,CDT2,CDT3,CDT3bis,CDT4,CDT5}.\\
To realize photonic AB caging, let us consider light transport in a periodic rhombic chain of coupled optical resonators or waveguides, see Fig.1(a). For the sake of definiteness, we will refer explicitly to light propagation in a waveguide array \cite{M2bis,M9}, however our analysis holds for coupled optical resonators as well \cite{M5}.
In the nearest-neighbor approximation, the lattice shows  three bands with the dispersion relations given by \cite{Cage2}
\begin{equation}
E_0=0 \; , \;\;\; E_{\pm}= \pm 2 \kappa \sqrt{1+\cos(q)}
\end{equation}
where $\kappa>0$ is the hopping rate between nearest-neighbor sites.
The existence of a flat band is simply due to the bipartite character of the chain, so that {\it partial} localization is possible for certain initial excitations even in the absence of the magnetic field \cite{Cage2}. 
In the presence of an artificial gauge field $\mathbf{A}(\mathbf{r})$, the hopping rate $\kappa$ between the two waveguides (or resonators) at spatial positions $\mathbf{r}_n$ and $\mathbf{r}_m$ acquires an additional phase 
$\int_{\mathbf{r}_n}^{\mathbf{r}_m} d \mathbf{r} \cdot \mathbf{A}$, and the band structure can be obtained after the Peierls' substitution $\kappa \rightarrow \kappa \exp \left( i \int_{\mathbf{r}_n}^{\mathbf{r}_m} d \mathbf{r} \cdot \mathbf{A} \right)$. This yields the modified dispersion relations \cite{Cage2}
\begin{equation}
E_0=0 \; , \; \; \; E_{\pm}= \pm 2 \kappa \sqrt{1+\cos (\gamma /2) \cos(q-\gamma/2)}
\end{equation}
where $\gamma \equiv \oint d \mathbf{r} \cdot \mathbf{A}$ is the field flux threading each plaquette (square) of the lattice; see Fig.1(a). The most striking
feature is that, for $\gamma= \pi$, the spectrum is made up of three 
{\it flat} bands; see Fig.1(b). The minimally extended 
eigenstates corresponding to the three flat bands at $\gamma= \pi$ are displayed in Fig.1(c). 
As discussed in Refs.\cite{Cage2,Cage4}, a
complete lock-in of any wave packet spreading in the lattice is thus realized (the so-called AB cage). In the optical setting of Fig.1(a), a synthetic magnetic
field can be realized for both coupled-cavities or waveguides following the schemes introduced in Refs.\cite{M5,M8bis}. In particular, for coupled waveguides the gauge field is realized by proper longitudinal modulation of the propagation constants of the waveguides \cite{M8bis} via an optical analogue of photon-assisted tunneling \cite{Cage8}. In the nearest-neighbor tight-binding approximation, transport of discretized light in the rhombic waveguide lattice of Fig.1(a) is described by the following equations

 \begin{figure}[htb]
\centerline{\includegraphics[width=8.6cm]{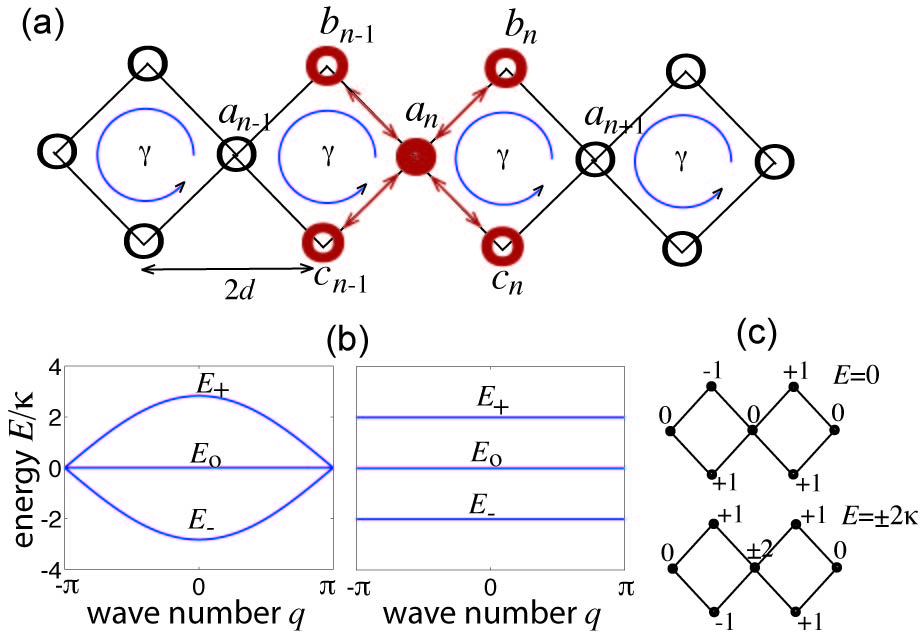}} \caption{ 
(Color online) Photonic caging in a rhombic lattice. (a) Rhombic chain of coupled optical waveguides. A synthetic magnetic flux $\gamma$ is applied in each plaquette. At $\gamma= \pi$ AB caging is realized: a photon initially localized on the central guide (filled circle) oscillates to its nearest neighbors (empty circles), but will not propagate through the lattice.
(c) Band structure of the lattice for $\gamma=0$ (left panel) and $\gamma= \pi$ (right panel), corresponding to band flattening and AB caging. (c) Localized eigenstates of energies $E_0=0$ and $E_{\pm}=2 \kappa$ for $\gamma= \pi$.}
\end{figure}

\begin{eqnarray}
i \frac{da_n}{dt} & = &  \kappa \left( b_n+b_{n-1}+c_{n}+c_{n-1} \right) + V_n(t) a_n \nonumber \\
i \frac{db_n}{dt} & = & \kappa \left( a_n + a_{n+1} \right)+W_n(t)b_n  \\
i \frac{dc_n}{dt} & = & \kappa \left( a_n + a_{n+1} \right)+X_n(t)b_n \nonumber
\end{eqnarray}
where $b_n$, $a_n$ and $c_n$ are the modal amplitudes in the upper, middle and lower rows of the chain, respectively, $t$ is the longitudinal spatial coordinate, and $V_n$, $W_n$, $X_n$ are the longitudinally-modulated propagation constants of the waveguides. After the gauge transformation $a_n=A_n \exp [-i \int_0^t dt' V_n(t')]$, $b_n=B_n \exp [-i \int_0^t dt' W_n(t')]$, $c_n=C_n \exp [-i \int_0^t dt' X_n(t')]$, Eqs.(3) take the form
\begin{eqnarray}
i \frac{dA_n}{dt} & = &  \kappa \left\{  B_n \exp(i \varphi_1)+B_{n-1} \exp(i \varphi_2) \right. \nonumber \\
& + & \left. C_{n} \exp(i \varphi_3)+C_{n-1} \exp(i \varphi_4) \right\}  \nonumber \\
\nonumber \\
i \frac{dB_n}{dt} & = & \kappa \left\{ A_n \exp(-i \varphi_1) + A_{n+1} \exp(-i \varphi_2) \right\}  \\
i \frac{dC_n}{dt} & = & \kappa \left\{  A_n \exp(-i \varphi_3) + A_{n+1} \exp(-i \varphi_4) \right\}  \nonumber
\end{eqnarray}
where we have set $\varphi_1(t)=\int_0^t dt'(V_n-W_n)$, $\varphi_2(t)=\int_0^t dt'(V_n-W_{n-1})$, $\varphi_3(t)=\int_0^t dt'(V_n-X_n)$ and $\varphi_4(t)=\int_0^t dt'(V_n-X_{n-1})$.
To realize a synthetic magnetic field, we slightly perturb the propagation constants $V_n$, $W_n$ and $X_n$ from the reference value $\beta_0$ by adding a stationary gradient term along the horizontal direction and rapidly-oscillating sinusoidal terms in the guides at the upper and lower rows, namely we assume
\begin{eqnarray}
V_n & =& \beta_0-2 \sigma n \nonumber \\
W_n & = & \beta_0 -(2n+1) \sigma+ A \cos (\omega t + \phi) \\
X_n & = & \beta_0 -(2n+1) \sigma -A \cos(\omega t-\phi) \nonumber
\end{eqnarray}  
where $\sigma$ is the linear gradient rate and $A$ is the amplitude of the sinusoidal modulation at spatial frequency $\omega$. In a practical design of the waveguide lattice,  the index gradient $\sigma$ is readily introduced by circularly bending the waveguides along the propagation direction $t$ \cite{CDT5}, whereas the weak sinusoidal modulation of the effective index of the guides in the upper and lower rows can be achieved by periodic modulation of the index change or waveguide width \cite{CDT3bis}. Note that the sinusoidal modulations in the upper and lower rows of the chain are out of phase by $ 2 \phi$. Note also that the phases $\varphi_l(t)$ ($l=1,2,3,4$) in Eqs.(4) are independent of the site index $n$. The spatial frequency $\omega$ and the gradient rate $\sigma$ are usually assumed to be much larger than the hopping rate $\kappa$. For $A=0$, evanescent tunneling among waveguides is inhibited owing to the large propagation constant mismatch between adjacent guides. 
Tunneling is restored in the presence of the longitudinal index modulation ($A \neq 0$) provided that the resonance condition $\sigma= M \omega$ is satisfied for some integer number $M$ \cite{M8bis}. In this case the phases $\varphi_l(t)$ in Eqs.(4) are periodic in $t$ with spatial period $T= 2 \pi / \omega$, and Floquet theory applies. The quasi-energy spectrum $\epsilon=\epsilon(q)$ of Eqs.(4) comprises three bands and can be numerically computed by standard methods \cite{Cage4} after introduction of the Ansatz $(A_n,B_n,C_n)^T=(A,B,C)^T \exp [iqn-i \epsilon(q)t]$, where the functions $A(t),B(t)$ and $C(t)$ are periodic with period $T$. For the sake of definiteness, the quasi energy spectrum is taken in the range $-\omega/2 \leq \epsilon < \omega/2$. As an example, Fig.2 shows the numerically-computed quasi energy spectrum as a function of the normalized amplitude modulation $\Gamma=A/\omega$ for a few values of the phase offset $2 \phi$ and for increasing values of the ratio $\omega / \kappa$. For $\phi=0$, i.e. for in-phase modulation, 
one quasi energy band is flat, corresponding to $\epsilon=0$, whereas the other two quasi energy bands show narrowing and pseudo-collapses at some special values of the normalized amplitude $\Gamma$; see Fig.2(a). As it will be discussed below, in the high-frequency regime such a pseudo-collapse is related to coherent destruction of tunneling. For an out-of-phase modulation, i.e. for $\phi \neq 0$, the three quasi energy bands are not flat and local narrowing and pseudo-collapses can be oserved like in the $\phi=0$ case; see Fig.2(b).   
 Interestingly, at $\phi=\pi/(4M)$ and in the high-frequency modulation limit the three quasi-energy bands collapse into three flat bands, {\it regardless of the value of the amplitude modulation} $\Gamma$; see Fig.2(c).  This regime corresponds to the AB caging, which is very distinct from coherent destruction of tunneling. To explain the behavior of the quasi energy bands as the phase $\phi$ is varied, let us consider the limit of a high frequency modulation $\omega \gg \kappa$. In this regime, the rapidly oscillating terms of  the couplings $\kappa \exp[ \pm i \varphi_l(t)]$ entering in Eqs.(4) can be disregarded and only the cycle-averaged terms $ \langle \kappa \exp[ \pm i \varphi_l(t)] \rangle$ can be kept at first approximation.  This yields the effective coupled-mode equations

 \begin{figure}[htb]
\centerline{\includegraphics[width=8.6cm]{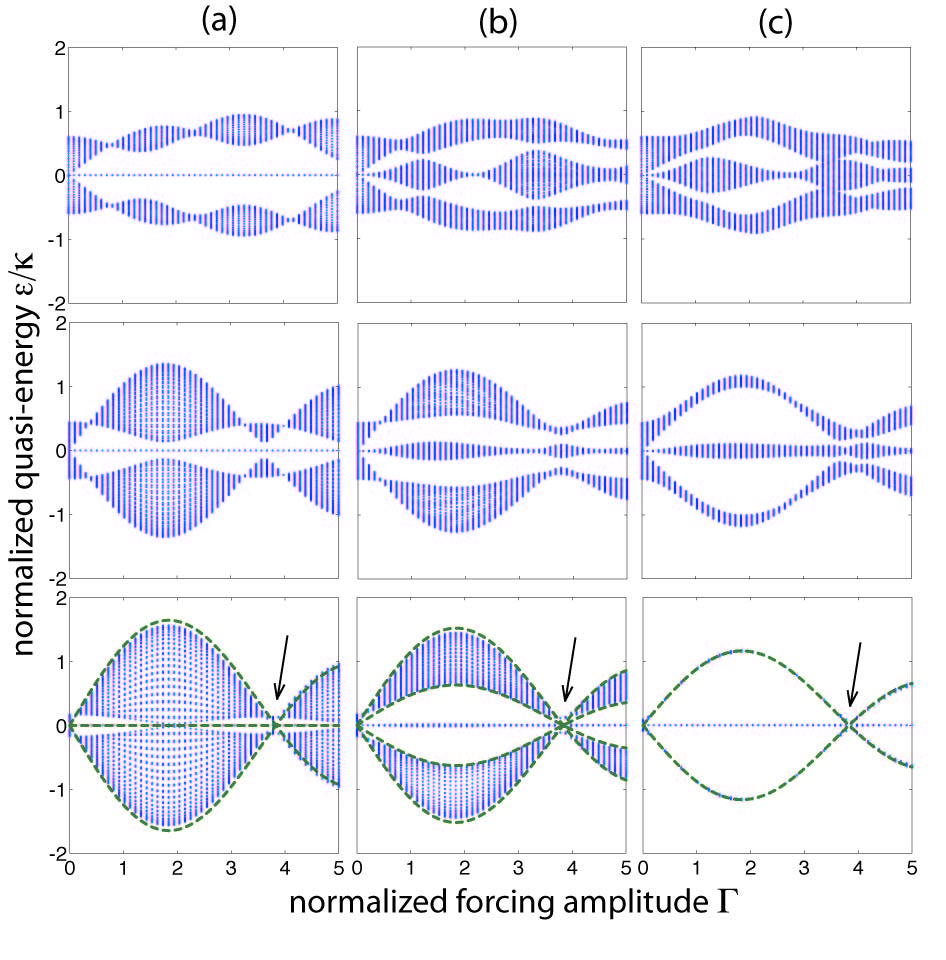}} \caption{Numerically-computed quasi-energy spectrum $\epsilon$ (in units of $\kappa$) versus the normalized forcing amplitude $\Gamma=A/ \omega$ of the modulated lattice [Eqs.(4)] for $M=1$ and for (a) $\phi=0$ (in-phase modulation), (b) $\phi=\pi/8$, and (c) $\phi=\pi/4$ (corresponding to AB caging). The upper, middle and lower panels refer to $\omega / \kappa=2$, $\omega / \kappa=5$, and $\omega / \kappa=15$, respectively. 
In the lower panels ($\omega / \kappa=15$) the dashed lines correspond the behavior of the quasi-energy spectrum as predicted in the high-frequency limit by the asymptotic analysis [Eqs.(6)], whereas the arrows show the forcing regime corresponding to coherent destruction of tunneling.}
\end{figure}

\begin{eqnarray}
i \frac{dA_n}{dt} & \simeq &  \kappa_0 \left\{  B_n \exp(i \phi_1)+  B_{n-1} \exp(i \phi_2) \right. \nonumber \\
& + & \left. C_{n} \exp(i \phi_1)+C_{n-1} \exp(i \phi_2) \right\}  \nonumber \\
\nonumber \\
i \frac{dB_n}{dt} & \simeq & \kappa_0 \left\{ A_n \exp(-i \phi_1) + A_{n+1} \exp(-i \phi_2) \right\}  \\
i \frac{dC_n}{dt} & \simeq & \kappa_0  \left\{  A_n \exp(-i \phi_1) + A_{n+1} \exp(-i \phi_2) \right\}  \nonumber
\end{eqnarray}
 where we have set $\kappa_0=\kappa \mathcal{J}_M(\Gamma)$,
 \begin{equation}
 \phi_1=M(\pi + \phi)+ \Gamma \sin \phi \; , \; \; \phi_2=-M \phi+\Gamma \sin \phi
 \end{equation}
 and $\mathcal{J}_M$ is the Bessel function of first kind and order $M$. The system (6) effectively realizes a rhombic lattice threaded by a magnetic field, where the phase $\phi$ of the sinusoidal modulation controls the effective magnetic flux $\gamma$ in each plaquette. In fact, it can be readily shown that the energy spectrum of Eqs.(6) is give by Eqs.(2), with $\kappa$ replaced by $\kappa_0=\kappa \mathcal{J}_M(\Gamma)$ and with
 \begin{equation}
 \gamma=2( \phi_1-\phi_2)=4 M \phi.
 \end{equation} 
 Note that, for a normalized modulation amplitude $\Gamma$ such that $\mathcal{J}_M(\Gamma)=0$ (e.g. $\Gamma \simeq 3.83$), one has $\kappa_0 = 0$ and thus coherent destruction of tunneling is realized, regardless of the value of the magnetic flux $\gamma$ (i.e. of $\phi$); see the arrows in the lower panels of Fig.2. This kind of trapping and the related phenomenon of dynamic localization have been discussed and demonstrated in several previous works \cite{CDT1,CDT2,CDT3,CDT3bis,CDT4,CDT5}. AB caging is a different form of localization which is observed for  $\phi=\pi/ (4 M)$, corresponding  to the effective magnetic flux $\gamma= \pi$. In this case the three energy bands are flat, regardless of the value of $\Gamma$; see the lower plot in Fig.2(c). An example of light trapping due to AB caging is shown in Fig.3. The figure shows the numerically computed evolution of the light intensity distributions $|a_n|^2$, $|b_n|^2$ and $|c_n|^2$ in a waveguide lattice for parameter values $\omega / \kappa=10$, $\sigma/ \omega =M=1$, $\Gamma=2$, and for two different values of $\phi$, $\phi=0$ (corresponding the absence of the magnetic field) and $\phi=\pi/4$ (corresponding to AB caging). The waveguide lattice is initially excited in a single waveguide belonging the central row. The figure clearly shows that light caging is realized for $\phi=\pi/4$, whereas ballistic transport (discrete diffraction) is observed for $\phi=0$. To get an idea of physical parameters corresponding to the simulations of Fig.3, let us consider a rhombic waveguide lattice manufactured in fused silica by femtosecond laser writing \cite{M3,CDT3bis} and probed at $\lambda=633$ nm. Assuming an horizontal waveguide spacing $2d=27 \; \mu$m between adjacent waveguides belonging to the same row [see Fig.1(a)], the typical coupling rate between nearest-neighbor guides is $\kappa \sim 1 \; {\rm cm}^{-1}$. The required index gradient $\sigma$ for the resonance condition $\omega=\sigma$ is $\sigma \simeq 10 \; {\rm cm}^{-1}$, which can be obtained by circularly-bending the waveguides in the horizontal plane with a radius of curvature \cite{CDT5} $R=2 \pi n_s d/(\sigma \lambda) \simeq 19.56 \;$cm, where $n_s\simeq 1.46$ is the substrate refractive index. The refractive index of the waveguides in the upper and lower rows is modulated along the longitudinal direction with a spatial modulation period $T= 2 \pi / \omega \simeq 6.28$ mm and modulation depth $\delta n \simeq \Gamma \lambda /T \simeq 2 \times 10^{-4}$. Such a modulation can be achieved by periodically varying the speed of the writing femtosecond laser beam, as demonstrated in Ref.\cite{CDT3bis}.  The total length of the waveguide array, corresponding to the maximum normalized propagation distance shown in Fig.3, is $t=10 / \kappa \simeq 10$ cm. By varying the phase offset $2\phi$ between the modulation of upper and lower waveguides,  from $2 \phi=0$ (in-phase modulation) to $2 \phi= \pi/2$ (quadrature-phase modulation), one should be able to observe a transition from ballistic to the AB caging regimes. Our numerical results did not account for losses arising from waveguide bending, which should be considered for a proper array design \cite{CDT5}. However, like in similar experiments on dynamic localization synthetic gauge fields in curved waveguide lattices \cite{M9,CDT4} radiation losses are not expected to destroy interference effects leading to AB caging. \par

 \begin{figure}[htb]
\centerline{\includegraphics[width=8.6cm]{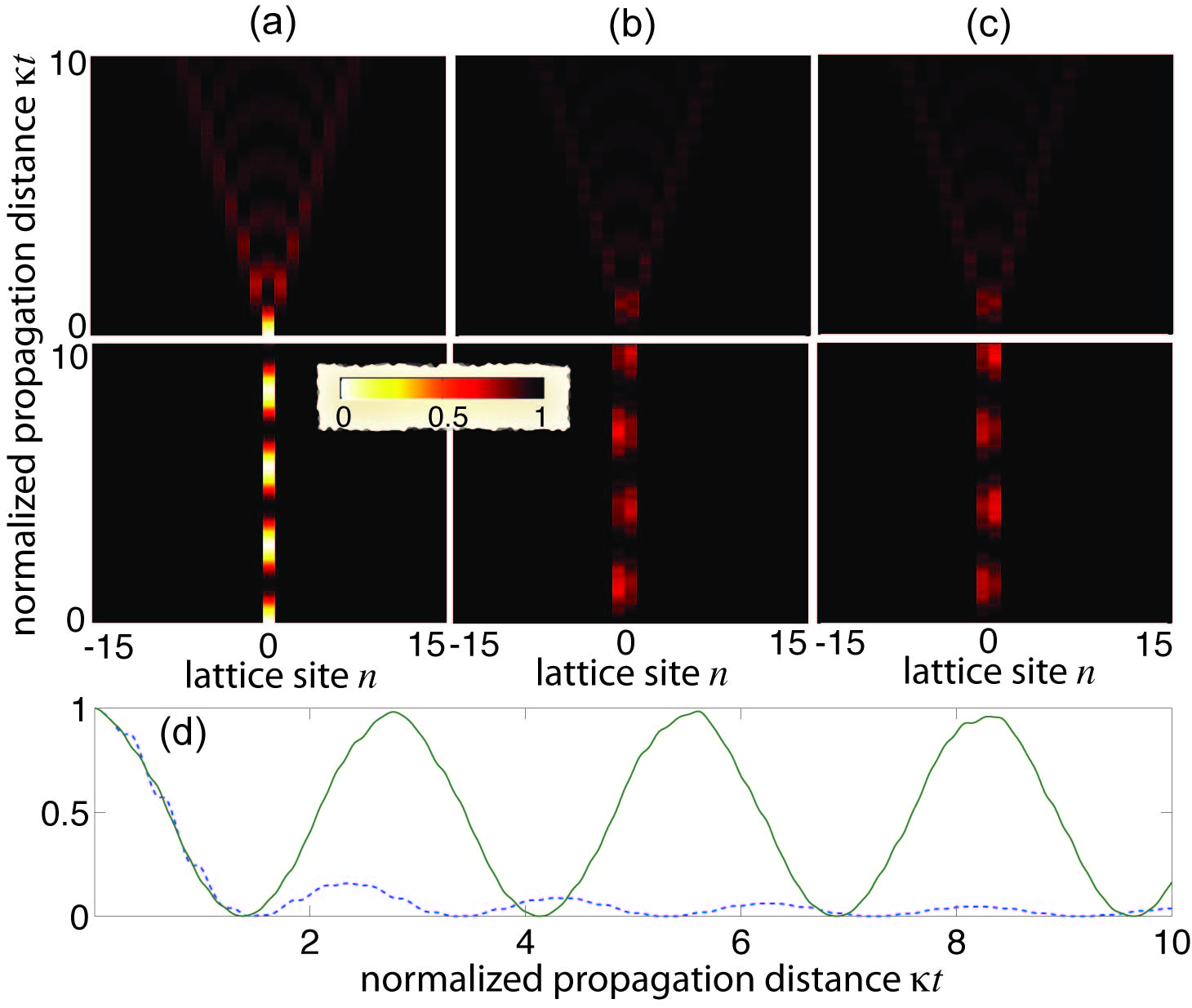}} \caption{(a-c) Numerically-computed evolution of light intensity distributions (pseudocolor maps) in the modulated lattice waveguides for parameter values $\omega/ \kappa=10$, $M=1$, $\Gamma=2$, and for $\phi=0$ (upper panels) and $\phi=\pi/4$ (lower panels). (a): snapshot of $|a_n|^2$; (b): snapshot of $|b_n|^2$; (c) snapshot of $|c_n|^2$. The lattice is excited at input plane in the waveguide of the middle row at site $n=0$.  In (d) the evolution of the normalized light intensity $|a_0|^2$, i.e. in the initially excited waveguide, is shown for $\phi=0$ (dashed curve) and for $\phi= \pi/4$ (solid curve). Photonic AB caging is clearly observed for $\phi= \pi/4$.}
\end{figure}

\par 
In conclusion, a mechanism of photon trapping in low-dimensional lattices of optical waveguides or coupled resonators has been proposed, which is based on an optical analogue of Aharonov-Bohm caging of charged particles in a magnetic field. This kind of localization requires a synthetic gauge field and it is thus rather different than other forms of trapping in ordered lattices, such as dynamic localization or coherent destruction of tunneling. Our study has been focused to ordered quasi one-dimensional rhombic lattices, however the analysis could be extended to consider other kinds of lattices, such as $\mathcal{T}_3$ tiling \cite{Cage1},  as well as the role of disorder and topological edge states in finite lattices \cite{Cage3,Cage8}.

\newpage

\footnotesize {\bf References with full titles}\\
\\
1. S. Raghu and F.D.M. Haldane, "Analogs of quantum-Hall-effect edge states in
photonic crystals", Phys. Rev. A {\bf 78}, 033834 (2008).\\
2. Z. Wang, Y. Chong, J.D. Joannopoulos,  and M. Soljacic,  "Reflection-free
one-way edge modes in a gyromagnetic photonic crystal", Phys. Rev. Lett.
{\bf 100}, 013905 (2008).\\
3. S. Longhi, "Bloch dynamics of light waves in helical optical waveguide arrays", Phys. Rev. B {\bf 76}, 19511 (2007).\\
4. Z. Wang, Y. Chong, J.D. Joannopoulos, and M. Soljacic, "Observation of
unidirectional backscattering-immune topological electromagnetic states",
Nature {\bf 461}, 772 (2009).\\
5.  M. Hafezi, E.A. Demler, M.D. Lukin, and J.M. Taylor, "Robust optical delay lines
with topological protection", Nature Phys. {\bf 7}, 907 (2011).\\
6. K. Fang, Z. Yu, and S. Fan, "Realizing effective magnetic field for photons by
controlling the phase of dynamic modulation", Nature Photon. {\bf 6}, 782 (2012).\\
7. V. Yannopapas,  "Topological photonic bands in two-dimensional networks of metamaterial elements ", New J. Phys. {\bf 14}, 11301 (2012).\\
8.  M.C. Rechtsman, J.M. Zeuner, A. T\"{u}nnermann, S. Nolte,	
M. Segev, and A. Szameit, "Strain-induced pseudomagnetic field and photonic Landau levels in dielectric structures", Nature Photon. {\bf 7}, 153 (2013).\\
9. L. Lu, L. Fu, J.D. Joannopoulos, and M. Soljacic, "Weyl points and line nodes in gyroid photonic crystals",  Nature Photon. {\bf 7}, 294 (2013).\\ 
10. M. Hafezi, S. Mittal, J. Fan, A. Migdall, and J. M. Taylor, "Imaging topological edge states in silicon photonics", Nature Photon. {\bf 7}, 1001 (2013).\\
11. S. Longhi, "Effective magnetic fields for photons in waveguide and coupled resonator lattices", Opt. Lett. {\bf 38}, 3570 (2013).\\
12. M. C. Rechtsman, J.M. Zeuner, Y. Plotnik, Y. Lumer, D. Podolsky, F. Dreisow, S.Nolte,M. Segev, and A. Szameit, "Photonic Flouet Topological Insulators",
Nature {\bf 496}, 196 (2013).\\
13. V.Yannopapas, "Gapless surface states in a lattice of coupled cavities: A photonic analog of topological crystalline insulators", Phys. Rev. B {\bf 84}, 195126 (2011).\\
14. S. Longhi, "Light transfer control and diffraction management in circular fibre waveguide arrays", J. Phys. B {\bf 40}, 4477 (2007).\\
15. M. Ornigotti, G. Della Valle, D. Gatti, and S. Longhi, "Topological suppression of optical tunneling in a twisted annular fiber", Phys. Rev. A {\bf 76}, 023833 (2007).\\
16. C.A. Dartoraa, K.Z. Nobregab, and G.G. Cabrerac, "Optical analogue of the AharonovÐBohm effect using anisotropic media", Phys. Lett. A {\bf 375}, 2254 (2011).\\ 
17. K. Fang, Z. Yu, and S. Fan, "Photonic Aharonov-Bohm effect based on dynamic modulation", Phys. Rev. Lett. {\bf 108}, 153901 (2012).\\
18. N. Satapathy, D. Pandey, P. Mehta, S. Sinha, J. Samuel, and H. Ramachandran, "Classical light analogue of the nonlocal Aharonov-Bohm effect", EPL {\bf 97}, 50011 (2012).\\
19. K. Fang, Z. Yu, and S. Fan, "Experimental demonstration of a photonic Aharonov-Bohm effect at radio frequencies", Phys. Rev. B {\bf 87}, 060301(2013).\\
20. E. Li, B.J. Eggleton, K. Fang, and S. Fan, "Photonic AharonovÐBohm effect in photonÐphonon interactions", Nature Commun. {\bf 5}, 3225 (2014).\\
21. J. Vidal, R. Mosseri, and B. Doucot, "Aharonov-Bohm Cages in Two-Dimensional Structures", Phys. Rev. Lett. {\bf 81}, 5888 (1998).\\
22. J. Vidal, B. Doucot, R. Mosseri, and P. Butaud, "Interaction Induced Delocalization for Two Particles in a Periodic Potential", Phys. Rev. Lett. {\bf 85}, 3906 (2000).\\
23. J. Vidal, P. Butaud, B. Doucot,  and R. Mosseri, "How to escape Aharonov-Bohm cages?", Phys. Rev. B {\bf 64}, 155306 (2001).\\
24. C.E. Creffield and G. Platero, "Coherent control of interacting particles using dynamical and Aharonov-Bohm phases", Phys. Rev. Lett. {\bf  105}, 086804  (2010).\\
25. C.C. Abilio, P. Butaud, Th. Fournier, B. Pannetier, J. Vidal, S. Tedesco, and B. Dalzotto, " Magnetic Field Induced Localization in a Two-Dimensional Superconducting Wire Network", Phys. Rev. Lett. {\bf 83}, 5102 (1999).\\
26. C. Naud, G. Faini, and D. Mailly, "Aharonov-Bohm Cages in 2D Normal Metal Networks ", Phys. Rev. Lett. {\bf 86}, 5104 (2001).\\
27. I.M. Pop, K. Hasselbach, O. Buisson, W. Guichard, B. Pannetier, and I. Protopopov,  "Measurement of the current-phase relation of Josephson junction rhombi chains", Phys. Rev. B {\bf 78}, 104504 (2008).\\
28.  A. Bermudez, T. Schaetz, and D. Porras,  "Synthetic Gauge Fields for Vibrational Excitations of Trapped ions", Phys. Rev. Lett. {\bf 107}, 150501 (2011).\\
29. S Longhi, M Marangoni, M Lobino, R Ramponi, P Laporta, E Cianci, and V Foglietti, "Observation of dynamic localization in periodically curved waveguide arrays", Phys. Rev. Lett. {\bf 96}, 243901 (2006).\\
30. R. Iyer, J.S. Aitchison, J. Wan, M.M. Dignam, and C.M. de Sterke, "Exact dynamic localization in curved AlGaAs optical waveguide arrays", Opt. Express {\bf 15}, 3212 (2007).\\ 
31. G Della Valle, M Ornigotti, E Cianci, V Foglietti, P Laporta, and S Longhi, "Visualization of coherent destruction of tunneling in an optical double well system", Phys. Rev. Lett. {\bf 98}, 263601 (2007).\\
32. A. Szameit, Y. V. Kartashov, F. Dreisow, M. Heinrich, T. Pertsch, S. Nolte, A. T\"{u}nnermann, V. A. Vysloukh, F. Lederer, and L. Torner, "Inhibition of Light Tunneling in Waveguide Arrays", Phys. Rev. Lett. {\bf 102}, 153901 (2009).\\
33. A. Szameit, I.L. Garanovich, M. Heinrich, A.A. Sukhorukov, F. Dreisow, T. Pertsch, S. Nolte, A. T\"{u}nnermann, S. Longhi, and Y.S. Kivshar,
"Observation of two-dimensional dynamic localization of light", Phys. Rev. Lett. {\bf 104}, 223903 (2010).\\
34. I.L. Garanovich, S. Longhi, A.A. Sukhorukov, and Y.S. Kivshar, "Light propagation and localization in modulated photonic lattices and waveguides",
Phys. Rep. {\bf 518}, 1 (2012).\\


\begin{thebibliography}{99}




\bibitem{M1}
S. Raghu and F.D.M. Haldane, Phys. Rev. A {\bf 78}, 033834 (2008).

\bibitem{M2}
Z. Wang, Y. Chong, J.D. Joannopoulos,  and M. Soljacic, Phys. Rev. Lett.
{\bf 100}, 013905 (2008).

\bibitem{M2bis}
S. Longhi, Phys. Rev. B {\bf 76}, 19511 (2007).

\bibitem{M3}
Z. Wang, Y. Chong, J.D. Joannopoulos, and M. Soljacic,  Nature {\bf 461}, 772 (2009).

\bibitem{M4}
M. Hafezi, E.A. Demler, M.D. Lukin, and J.M. Taylor, Nature Phys. {\bf 7}, 907 (2011).

\bibitem{M5}
K. Fang, Z. Yu, and S. Fan, Nature Photon. {\bf 6}, 782Ð787 (2012).

\bibitem{ref2}
V. Yannopapas,  New J. Phys. {\bf 14}, 11301 (2012).

\bibitem{M6} 
M.C. Rechtsman, J.M. Zeuner, A. T\"{u}nnermann, S. Nolte,	
M. Segev, and A. Szameit,  Nature Photon. {\bf 7}, 153 (2013).

\bibitem{M7}
L. Lu, L. Fu, J.D. Joannopoulos, and M. Soljacic, Nature Photon. {\bf 7}, 294 (2013).

\bibitem{M8}
M. Hafezi, S. Mittal, J. Fan, A. Migdall, and J. M. Taylor, Nature Photon. {\bf 7}, 1001 (2013).

\bibitem{M8bis}
S. Longhi, Opt. Lett. {\bf 38}, 3570 (2013).

\bibitem{M9}
 M. C. Rechtsman, J.M. Zeuner, Y. Plotnik, Y. Lumer, D. Podolsky, F. Dreisow, S.Nolte,M. Segev, and A. Szameit,
Nature {\bf 496}, 196 (2013).

\bibitem{ref1}
V.Yannopapas, Phys. Rev. B {\bf 84}, 195126 (2011).

\bibitem{AB0}
S. Longhi, J. Phys. B {\bf 40}, 4477 (2007).

\bibitem{AB1}
M. Ornigotti, G. Della Valle, D. Gatti, and S. Longhi, Phys. Rev. A {\bf 76}, 023833 (2007).

\bibitem{AB2}
C.A. Dartoraa, K.Z. Nobregab, and G.G. Cabrerac, Phys. Lett. A {\bf 375}, 2254 (2011). 

\bibitem{AB3}
 K. Fang, Z. Yu, and S. Fan, Phys. Rev. Lett. {\bf 108}, 153901 (2012).

\bibitem{AB4}
N. Satapathy, D. Pandey, P. Mehta, S. Sinha, J. Samuel, and H. Ramachandran, EPL {\bf 97}, 50011 (2012).

\bibitem{AB5}
K. Fang, Z. Yu, and S. Fan, Phys. Rev. B {\bf 87}, 060301(2013).

\bibitem{AB6}
E. Li, B.J. Eggleton, K. Fang, and S. Fan, Nature Commun. {\bf 5}, 3225 (2014).

\bibitem{Cage1}
J. Vidal, R. Mosseri, and B. Doucot, Phys. Rev. Lett. {\bf 81}, 5888 (1998).

\bibitem{Cage2}
J. Vidal, B. Doucot, R. Mosseri, and P. Butaud, Phys. Rev. Lett. {\bf 85}, 3906 (2000).

\bibitem{Cage3}
J. Vidal, P. Butaud, B. Doucot,  and R. Mosseri, Phys. Rev. B {\bf 64}, 155306 (2001)

\bibitem{Cage4}
C.E. Creffield and G. Platero, Phys. Rev. Lett. {\bf  105}, 086804  (2010).

\bibitem{Cage5}
C.C. Abilio, P. Butaud, Th. Fournier, B. Pannetier,
J. Vidal, S. Tedesco, and B. Dalzotto, Phys. Rev. Lett. {\bf 83}, 5102 (1999)

\bibitem{Cage6}
C. Naud, G. Faini, and D. Mailly, Phys. Rev. Lett. {\bf 86}, 5104 (2001).

\bibitem{Cage7}
 I.M. Pop, K. Hasselbach, O. Buisson, W. Guichard, B. Pannetier, and I. Protopopov, Phys. Rev. B {\bf 78}, 104504 (2008).
 
 \bibitem{Cage8}
 A. Bermudez, T. Schaetz, and D. Porras, Phys. Rev. Lett. {\bf 107}, 150501 (2011).

\bibitem{CDT1}
S Longhi, M Marangoni, M Lobino, R Ramponi, P Laporta, E Cianci, and V Foglietti, Phys. Rev. Lett. {\bf 96}, 243901 (2006).

\bibitem{CDT2}
R. Iyer, J.S. Aitchison, J. Wan, M.M. Dignam, and C.M. de Sterke, Opt. Express {\bf 15}, 3212 (2007). 

\bibitem{CDT3}
 G Della Valle, M Ornigotti, E Cianci, V Foglietti, P Laporta, and S Longhi, Phys. Rev. Lett. {\bf 98}, 263601 (2007).
 
 \bibitem{CDT3bis}
 A. Szameit, Y. V. Kartashov, F. Dreisow, M. Heinrich, T. Pertsch, S. Nolte, A. T\"{u}nnermann, V. A. Vysloukh, F. Lederer, and L. Torner, Phys. Rev. Lett. {\bf 102}, 153901 (2009).

\bibitem{CDT4}
 A. Szameit, I.L. Garanovich, M. Heinrich, A.A. Sukhorukov, F. Dreisow, T. Pertsch, S. Nolte, A. T\"{u}nnermann, S. Longhi, and Y.S. Kivshar, Phys. Rev. Lett. {\bf 104}, 223903 (2010).

\bibitem{CDT5}
I.L. Garanovich, S. Longhi, A.A. Sukhorukov, and Y.S. Kivshar, Phys. Rep. {\bf 518}, 1 (2012).

















\end{thebibliography}
\end{document}